\documentclass[12pt]{article}

\usepackage{subfigure}
\usepackage{epsfig}
\usepackage{epsf}
\usepackage{graphics}
\usepackage{graphicx}

\def\arcdeg{\hbox{$^\circ$}}

\def\arcpt{${{\lower3pt\hbox{$^{\prime\prime}$}}\atop{\raise4pt\hbox{.}}}$}
\newcommand\etal{\mbox{\textit{et al.~}}}

\begin{document}
\begin{center}
\huge
{\bf Astrometry - Challenging our Understanding of Stellar Structure and Evolution}
\normalsize
\vskip10pt
\end{center}

\begin{center}
G. Fritz Benedict (U. Texas, fritz@astro.as.utexas.edu),Todd J. Henry (GSU)\\Rob Olling (U MD)
\end{center}

\begin{figure}[h]
\centering
\includegraphics[scale=0.45, angle=0]{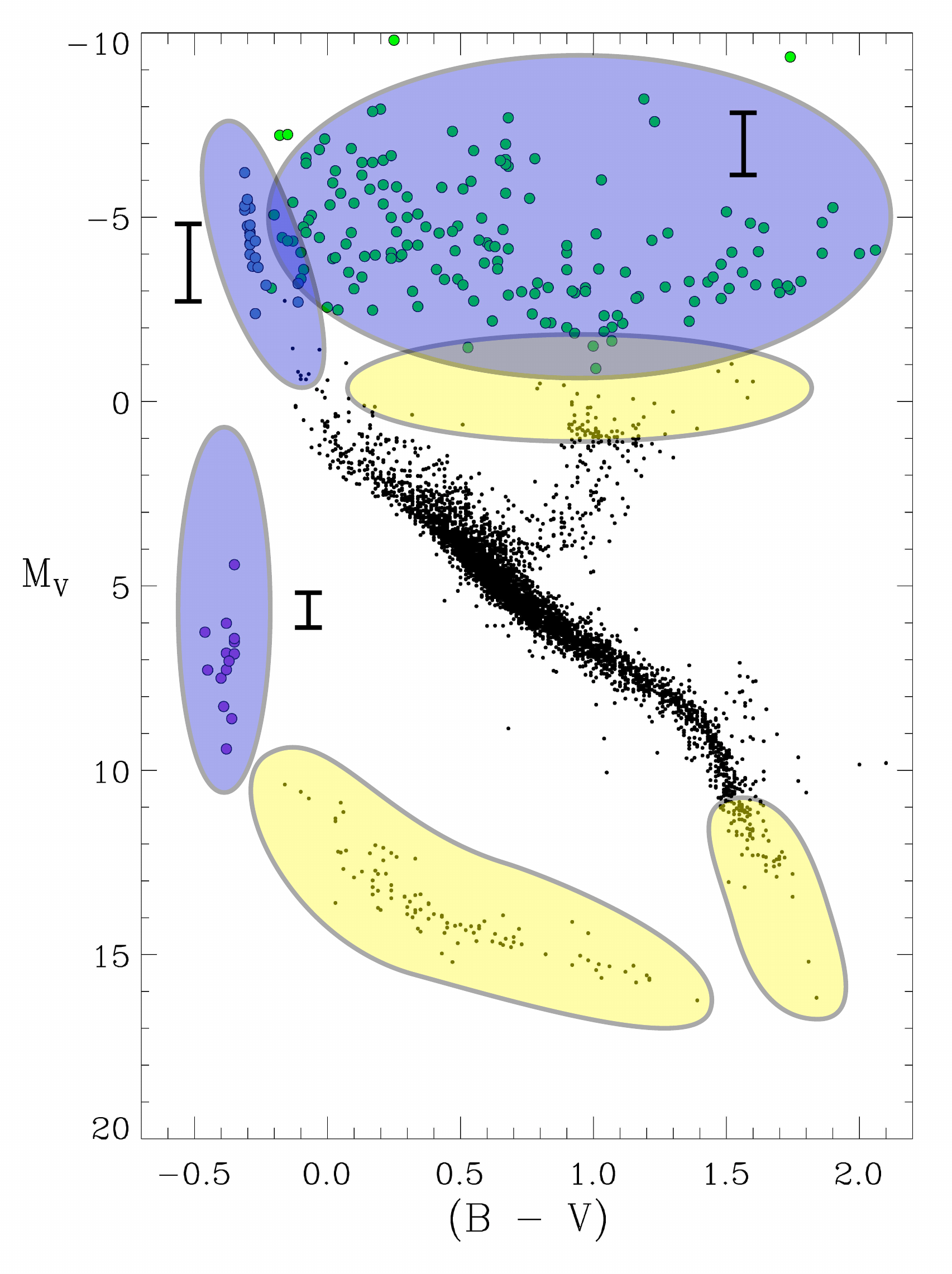}
\end{figure}
\begin{center}
White Paper Prepared for the 2010 Decadal Survey\\ Stars and Stellar Evolution Panel\\ Galactic Neighborhood Panel\\7 February 2009
\end{center}

\clearpage

\begin{center}
\Large
{\bf ABSTRACT}
\end{center}
\normalsize

Stellar mass plays a central role in our understanding of star formation and aging.  Stellar astronomy is largely based on two maps, both dependent on mass, either indirectly or directly: the Hertzprung-Russell Diagram (HRD) and the Mass-Luminosity Relation (MLR).  
The extremes of both maps, while not {\it terra incognita}, are characterized by large uncertainties. A precise HRD requires precise distance obtained by direct measurement of parallax. A precise MLR requires precise measurement of binary orbital parameters, with the ultimate goal the critical test
of theoretical stellar models. Such tests require mass accuracies of $\sim$1\%. Substantial improvement in both maps requires astrometry with microsecond of arc measurement precision. Why? First, the 'tops' of both stellar maps contain relatively rare objects, for which large populations are not found until the observing horizon reaches hundreds or thousands of parsecs. Second, the 'bottoms' and 'sides' of both maps contain stars, either intrinsically faint,  or whose rarity guarantees great distance, hence apparent faintness.   With an extensive collection of high accuracy masses that can only be provided by astrometry with microsecond of arc measurement precision, astronomers will be able to stress test theoretical models of stars at any mass and at every stage in their aging processes.

\section {\bf Distances to Objects at the Extremes of the Hertzsprung-Russell Diagram}

The HRD is generally the first figure a stellar astronomer considers
to understand any given star in context.  All  classes of
stars appear on the HRD, including supergiants, AGB stars, giants,
subgiants, dwarfs, subdwarfs, and white dwarfs, as well as other
exotic stars.  The HRD maps a star's temperature and luminosity, which
together determine the star's radius.  However, placing a star on the
HR diagram requires knowledge of its luminosity, thus an accurate
distance measurement.  Trigonometric parallax is the most reliable and
straightforward method of measuring stellar distances, and is usually
the most accurate method as well.  Ground-based parallax efforts have
pushed forward for 170 years (e.g., the summary in the Yale
Parallax Catalog, van Altena \etal 1995). In the past 20 years, space-based
efforts have made great headway, from Hipparcos results (ESA 1997; van
Leeuwen 2008) to the Hubble Space Telescope (Benedict \etal 2007). Figure 1 shows an observational HRD. 
In the coming era, astrometric efforts like Pan-STARRS, LSST, and
Gaia will measure parallaxes of millions of stars to unprecedented
precision.  Even so, there remain rare, astrophysically
compelling objects  at such great distances in the Galaxy that only astrometers with precision better than 10 microseconds of arc (10 $\mu$as) and heretofore unprecedented sensitivity can provide accurate luminosities.


$\bullet$The massive {\bf O stars}
are among the brightest objects observed in
galaxies and they play a central role in sculpting the ISM (through
their radiative and mechanical energy input), while driving the
chemical enrichment of galaxies.  With lives of only a few million
years, they quickly burn through their fuel and explode
catastrophically in supernovae.  However, the fundamental parameters
of these extraordinarily rare stars are still poorly known because
they are generally found at large distances.  Some O stars are found
in clusters, but roughly 20\% are runaways or field O stars while many
others are found in loose associations with poorly-defined boundaries
and distances, e.g., the Cep OB6 association has a 3$^\circ$ extent
and a consequent 5\% dispersion (1$\sigma$) in distance (Benedict et
al 2002).

The placement of massive stars in the HRD relies heavily at present on
model atmospheres.  Hot stars
all have essentially the same colors in the optical/IR spectral range
(after correcting for interstellar reddening), so estimates of their
effective temperatures are made by comparing spectral line profiles
with those calculated from sophisticated models (Repolust et al 2004);
the resulting temperature estimates are typically accurate only to 5\%.
Their luminosities are determined from their absolute magnitudes and
bolometric corrections (again derived from models for the estimated
temperature), but reliable absolute magnitudes are only available for
O stars in clusters where distances are known from other techniques.
For the majority of O stars, the absolute magnitudes are estimated
from spectral classification calibrations (based upon those stars in
clusters) which typically result in 25\% distance errors, consequently
resulting in luminosity errors approaching 50\%.  Thus, the
observational HRD only loosely constrains modern evolutionary models
for massive stars (Herrero et al 2007).
Progress will clearly require better distance measurements from
accurate trigonometric parallaxes.  


Distances accurate to 1\%  will provide the
accurate luminosities crucial to testing assumptions about
interior structure, in particular the role of rotation and meridional
circulation (Ekstr\"{o}m et al 2008), important for the kinds
of supernovae and compact remnants produced by massive stars.  A 1\%
error also corresponds to the typical errors in luminosity from errors
in effective temperature that will come from new ground based
interferometric observations, and errors in angular size derived from
the spectral energy distribution and errors in extinction (Fitzpatrick
\& Massa 2007).  


$\bullet$ Found among the naked
eye stars are many famous {\bf supergiants}, many of which are not in
clusters, so fainter stars cannot be used as proxies for determining
distances.  Parallaxes accurate to 4 $\mu$as 
will enable astronomers to (1) pinpoint supergiants' luminosities
on the HRD, (2) understand how metallicities affect their positions,
and (3) improve the wind-momentum-luminosity  and
flux-weighted gravity-relations used to derive extragalactic
distances.  In addition, a bright star parallax program 
offers a valuable public outreach opportunity  --- astronomers will be able to tell anyone who might ask
where the stars they can see are in our Galaxy.

$\bullet$ Distances to {\bf Planetary Nebulae (PNe)} are important for understanding
the physics of the nebulae, the evolutionary state of the central
stars (e.g., time since the ejection of material), and the space
density and formation rate of PNe.  At present, however, distances are
notoriously uncertain, both in terms of systematic effects and for
individual nebulae.  Only 16 have measured trigonometric parallaxes
(Benedict et al 2003; Harris et al 2007), and distances are large
enough for most PNe 
to preclude many more being measured
until we achieve $\mu$as astrometric precision.
There are roughly 2000 PNe known in the Galaxy,
many in and around the Galactic bulge 
They include a large variety of types, and
understanding these different types adds scientific importance to
getting accurate distances to many PNe.  
One additional product of $\mu$as astrometry will be the
identification of binary motion for PNe central stars.  One theory of
the origin of bipolar symmetry seen in many PNe argues that binary
central stars are common.  A high-cadence $\mu$as campaign on a selected sample of bright (V$\sim$15) central
stars in bipolar PNe could
provide constraints on the frequency of binaries and characterize their orbits.
\section{\bf An Improved Mass-Luminosity Relation and Masses of Stars at Evolutionary Extremes}
\normalsize

Mass is arguably the single most important characteristic of a star,
as it determines a star's size and color, as well as how long it will
live and what fuels it will burn.  Knowing the masses of main sequence
stars answers basic astrophysical questions such as, {\it What is the
biggest star?  What is the smallest star?  How is the mass of a
stellar nursery partitioned into various types of stars?}  and, {\it
What is the mass content of the Galaxy and how does it evolve?}  To
answer these and other fundamental questions requires masses to 1\% accuracy.  Why 1\%? 
Our knowledge of stars consists of surface temperature, T$_{e}$; apparent magnitude; metallicity; distance, hence luminosity; and through T$_{e}$ (or long-baseline interferometry), radius; and stellar mass, $M$.
At a 5\% level of mass precision, luminosities are uncertain by 12 to 22\%.
This luminosity uncertainty means, for example, that radii would be very poorly determined, rendering them far less useful as checks of stellar models. At the 1\% level of mass precision the variation in luminosity is now only 2 to 4\%. This precision of luminosity will allow choices to be made between various modeling approaches, which could include stellar phenomena such as convection, mass loss, turbulent mixing, rotation, and magnetic activity (Andersen 1998). Of the
$\sim$40 stars with masses this accurately known (all in eclipsing
binaries), three-quarters have masses between 1 and 3 M$_\odot$, a limited range over which to test stellar models. 

The MLR's broad appeal is its applicability to many areas of
astronomy.  A reliable MLR lets us use a star's luminosity as a proxy
for its mass, which is a valuable commodity in radial velocity,
astrometric, cataclysmic binary, and extrasolar planet work.  In the
broader Galactic context, an accurate MLR provides benchmarks for
comparisons to objects in stellar clusters, and allows us to estimate
just how much of the ``missing'' mass is made up of the smallest
stars.  At the faint end of the stellar main sequence, the MLR is
crucial for brown dwarf studies because measurement of a sufficiently
small mass can demonstrate that a star is a {\it bona fide} brown
dwarf.  High accuracy masses are needed because (as shown in Figure 1)
the width of the main sequence on the MLR is 20\% or more at a given
luminosity.  Even though the individual stellar masses
calculated to date are determined to 5\% or better, they are of mixed
pedigree in age and metallicity.  An astrometer with $\mu$as precision
is needed to measure
masses accurate to 1\% in myriad environments and for a suite of
different kinds of stars, and stages of stellar evolution which until now have never resided in an MLR.

\begin{figure}
\includegraphics[scale=0.8, angle=0]{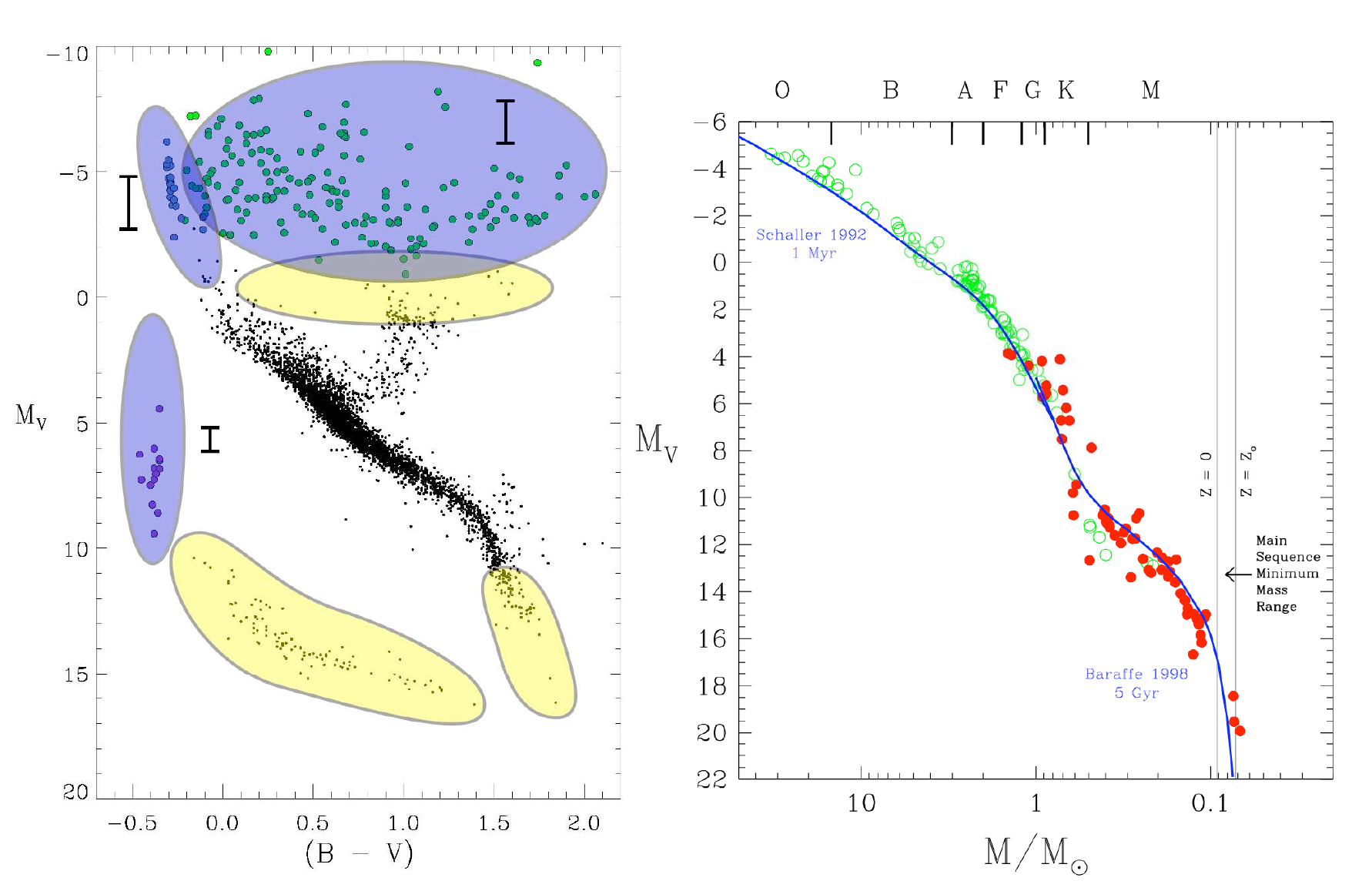}
\caption{{\bf Left:} Characteristic regions of the HR Diagram to be explored
effectively by 
astrometers with $\sigma_{\pi}<5 \mu$as (blue) and   $\sigma_{\pi}<10 \mu$as (yellow), with error bars
representative of current knowledge near the blue regions.  Supergiants
are from Hipparcos, with mean parallax error 37\%.  A representative sample of O stars is shown,
although only six parallaxes are currently available with errors less
than 20\%.  PNe central stars with
parallaxes from Harris et al (2007) are also shown, for which the
average parallax error is 19\%.  
Main sequence and
giant branch data points are from Hipparcos (van Leeuwen
2007 updated to 2008) 
Data for white dwarfs have been taken from
Bergeron et al (2001). {\bf Right:} The mass-luminosity relation in 2008, using eclipsing binary
data (open green points) from Andersen (1991) and others, supplemented
with visual binary data (solid red points) from Henry (private communication).  Model curves for the MLR at the
indicated ages and solar metallicity are shown, from Schaller et al
(1992) at high masses and Baraffe et al (1998) at low masses.  Note
the spread in empirical mass determinations at a given luminosity
throughout the main sequence, caused primarily by different ages and
metallicities.}
\end{figure}

Typically, stellar mass estimates come from measurements of the
orbital motions of binary stars.  The most accurate stellar
masses have been inferred from studies of eclipsing spectroscopic
binaries (Andersen 1991), but this method is limited.  At the highest masses, only a few
known eclipsing systems contain O stars (Gies 2003), and many of these
are interacting systems whose members may not be representative of
single stars.  At the lowest masses, stars are small, so few binaries
eclipse and visual binaries must be used (Henry et al 1999).  Other
rare, but important types of evolved stars remain almost completely
unmeasured.
To significantly improve the MLR at its extremes we require a combination of
exquisitely accurate astrometry, faint magnitude limit, bright magnitude limit, and flexible
scheduling. 

The 'gold standard' method, applicable no matter what the binary geometry,
consists of resolving the system and measuring the
relative orbits referenced to a grid of reference stars.  
Resolution is particularly important for the MLR because
the component luminosities must be measured to place the stars on the
MLR (and on the HRD).  To reach 1\% mass accuracy, an
inclination precision of 0.2\% is required for an orbit with $i = 45$\arcdeg,
assuming uncertainties in other orbital parameters do not dominate.
For resolvable binaries, the minimum requirement for a 1\% mass
determination is a 0.33\% distance measurement, which corresponds to
833 pc for a 4 $\mu$as measurement.  

$\bullet$ {\bf White dwarf} research has far reaching implications in diverse astronomical
fields, from cosmology to Galactic halo populations to nearby star
studies.  
From the youngest WDs found as central stars in PNe to the oldest WDs
from the halo, only $\mu$as faint-star astrometry brings unusual objects (with various compositions and ages) into reach for mass
determinations.
Nearly every aspect of WD research relies on the theoretical
mass-radius relation for WDs.  This relationship depends on the
internal composition of the WD.  Ideally, WD masses need to be
known to 1\% (or better) to stress test the mass-radius relation to
reveal the true chemical makeup of WDs, and permit us to discriminate,
for example, between different hydrogen envelope masses (Jordan 2007).

To date, empirical masses to support the theoretical mass-radius
relationship are severely limited --- only three WDs have dynamical
mass measurements known to better than 5\%, Sirius B, Procyon B, and
40 Eri B (Provencal \etal 2002).  Other WDs with masses, such as the
remaining 18 WDs that populate the mass-radius relation in Figure 13
of Provencal et al (2002), have masses gleaned from gravitational
redshift studies of common proper motion systems (in which a companion
is used to determine systemic parameters) or have spectroscopically
inferred masses.  In general, such mass determinations are rather
poorly constrained, with errors of 10\% or more.
Through ongoing studies of double degenerate systems with Hubble Space
Telescope's Fine Guidance Sensors, eight systems have already been
identified that might provide masses accurate to 1\% when examined
by an astrometer with $\mu$as precision.  At separations of a few to tens of milliarcseconds, these
systems can be resolved only by
long-baseline interferometry.  Such a device will play a crucial role in populating the
mass-radius diagram with multiple empirical checks of the theoretical
mass-radius relation while providing details of the internal
structures of WDs.  

$\bullet$ Although extremely {\bf massive stars} (O stars) are rare, as a group they are known to contain many
binaries, with a multiplicity fraction of 75\% for those found in
clusters or associations (Mason et al 1998).  
However, mass estimates
for such systems may tell us more about the evolutionary mass exchange
histories of binaries rather than providing fundamental data to
calibrate the properties of stars in general.  For non-interacting,
non-eclipsing O star binaries, masses are determined by supplementing
an SB2 orbit with a precisely determined orbital inclination, or by
resolving the binary and finding the shape of the orbit.  Because O
stars are rare and consequently distant, such measurements will
require $\mu$as astrometric  measurements.
For example 
HD 93205 resides in the Carina Nebula region at a distance of 2.6
kpc.  This is a 6d binary consisting of O3 V and O8 V stars
(Antokhina et al 2000), and the maximum photocenter motion will be in
the range of 9--16 $\mu$as (depending on the adopted flux ratio).  
Microsecond of arc measurement precision and flexible scheduling 
are required to obtain the orbital inclination and distance
of this system and to obtain the mass of a star at the top of the main
sequence.

$\bullet$  {\bf Pre-Main Sequence Stars}.
With the exception of solar mass objects, evolution of stars from
birth to the zero-age main sequence is poorly calibrated (Schaefer \etal
 2008 and references therein).  Binaries in star formation regions
provide an opportunity to determine precise dynamical masses in
low-mass, young star systems (e.g., Prato \etal 2002; Hillenbrand and
White 2004).  Fewer than 100 pre-main sequence (PMS) spectroscopic
binaries (SBs) are currently known (Melo \etal 2001), and even fewer
eclipsing PMS SBs have been identified (Stassun \etal 2007).  Mass
determinations  of a few dozen binaries among the youngest T
Tauri star populations that are accurate to a few percent would
revolutionize models of young star evolution.
Astrometric capability at the 4 $\mu$as level is required to accurately determine
the orbit of the photocenter of the shortest period ($<$100 days) T
Tauri SBs, providing the system inclinations and hence absolute
component masses to the required few percent precision for meaningful
calibration of evolutionary tracks. 

$\bullet$ {\bf Open star clusters} are laboratories for the study of stellar
astrophysics because they provide large numbers of stars with the same
age and chemical composition.  
To date, the only cluster for which an MLR
has been determined is the Hyades (Torres \etal 1997).  However, the
Hyades MLR extends only from 2.4 to 0.8 M$_\odot$ with mass errors of
5--10\%.  This MLR is insufficient for critical tests of the models
and does not include the smallest stars, for which the age and
metallicity effects are most pronounced.  An astrometer with $\mu$as accuracy is
needed to reduce these errors to the 1\% level needed for meaningful
analyses.
To maintain a 1\% mass accuracy beyond about 200 pc requires raw resolution and an astrometer that can be flexibly scheduled  to provide
coverage near crucial periastron passages to yield the most accurate
masses possible.  Of special interest is the production of a
reliable MLR for ancient (and distant!)  M67, whose constituent stars are all the same
age an metallicity as the Sun.  Overall, what is particularly
compelling  is that for a suitable astrometer target clusters span a range of 1000
in age, thereby providing a framework within which to study
many aspects of stellar evolution, once accurate distances and masses
are available.

\section{\bf The Coming Era of Microsecond of arc Astrometry: Gaia and SIM}
Both the Gaia and the Space Interferometry Mission (SIM) efforts will revolutionize our understanding of
stellar astrophysics via the HRD and MLR maps, albeit in different
ways.  Gaia's high precision astrometry of one billion sources will
provide superb measurements of luminosities, temperatures, and masses
of most of the stellar main sequence, giants, subdwarfs, and white
dwarfs.  More specifically, Gaia will determine distances to 1\% for
10$^7$ stars having V=6--13 within $\sim$1 kpc (Lindegren \etal 2008).

SIM provides complementary depth to Gaia's astrometry in specific
regimes of both magnitude and distance.  SIM can effectively observe
stars with V$\sim$ $-$1 to at least 18, adding complementary phase
space at bright magnitudes to Gaia's bright cutoff at V$\sim$6 and
making more accurate astrometric measurements at the faint end.  Thus,
only SIM can pinpoint the locations of many of famous naked-eye stars
in the night sky while opening up new territory for intrinsically
faint stars at tens or hundreds of parsecs.  For magnitudes 6--13,
SIM's wide-angle mode parallax precision of 4 $\mu$as is modestly
better than Gaia's 8 $\mu$as, which will observe a far larger stellar
sample.  For magnitudes 14--18, SIM's precision is 3--20 times better
than Gaia's (Lindegren \etal 2008).  
The combination of the ability to observe bright objects at all, and
faint objects with superior precision, provides several niches
important to stellar astronomy that only SIM can explore.  In nearly every category Gaia will be a pathfinder for SIM, much like the Palomar Schmidt was for the 200-inch telescope.

This white paper is based on Chapter 8, "SIM-Lite Astrometric Observatory", by T.  Henry , D. Gies, W. Jao, A. Riedel, J. Subasavage (GSU), G.  Benedict (U Texas), H. Harris (USNO), P. Ianna (U Virginia), J. Thorstensen (Dartmouth), C. Beichman (NExScI), L. Prato (Lowell 
Obs.), and M. Simon (SUNY Stony Brook).

\vskip20pt
{\bf References}
\normalsize
\tiny

Andersen, J. 1991, A\&A Reviews, 3, 91

Antokhina, E.A., \etal. 2000, ApJ, 529, 463

Baraffe, I.,\etal 1998, A\&A, 337, 403

Benedict, G.F., \etal 2002, AJ, 124, 1695

Benedict, G.F., \etal 2003, AJ, 126, 2549

Benedict, G.F., \etal 2007, AJ, 133, 1810

Bergeron, P., Leggett, S.K., \& Ruiz, M.T. 2001, ApJS, 133, 413


De Becker, M., Rauw, G., \& Manfroid, J. 2004, A\&A, 424, L39

ESA 1997, VizieR Online Data Catalog, 1239, 0

Ekstr\"{o}m, S., \etal. 2008, A\&A, 478, 467

Fitzpatrick, E.L., \& Massa, D. 2007, ApJ, 663, 320

Gies, D.R. 2003, Proc. IAU Symp. 212, ed. K. van der Hucht, A. Herrero, \&
Esteban, C.  (San Francisco: ASP), 91



Harris, H.C. \etal 2007, AJ, 133, 631

Henry, T.J. \etal 2006, AJ, 132, 2360

Henry, T.J. \etal 1999, ApJ, 512, 864

Herrero, A., \etal,  2007, ASP Conf. Ser. 367, ed. N. St-Louis \& A. F. J. Moffat (San Francisco: ASP), 67

Hillenbrand, L.A. \& White, R.J. 2004, ApJ, 604, 741

Jordan, S. 2007, 15th European Workshop on White Dwarfs, 372, 169

Lindegren, L. \etal 2008, IAU Symposium 248, 217

Mason, B.D., \etal 1998, AJ, 115, 821


Melo, C.H.F.,\etal 2001, A\&A, 378, 898



Prato, L., \etal. 2002, ApJ, 569, 863

Provencal, J.L., \etal 2002, ApJ, 568, 324

Repolust, T., Puls, J., \& Herrero, A. 2004, A\&A, 415, 349

Schaefer, G.H. \etal 2008, AJ, 135, 1659

Schaller, G., \etal 1992, A\$AS, 96, 269



Stassun, K.G., Mathieu, R.D.,\& Valenti, J.A. 2007, ApJ, 664, 1154


Torres, G., Stefanik, R.P. \& Latham, D.W. 1997, ApJ, 485, 167

van Altena, W.F., Lee, J.T. \& Hoffleit, E.D. 1995, Catalogue of Trigonometric Parallaxes

van Leeuwen, F. 2007, Hipparcos, the New Reduction of the Raw Data,
Astrophysics \& Space Sciences Library, 350, Springer
\end{document}